# Characterizing Aperture Masking Interferometry in the Near-Infrared as an Effective Technique for Astronomical Imaging

Kyle Morgenstein

MIT Class of 2020

Department of Aeronautical and Astronautical Engineering

Department of Earth Atmospheric and Planetary Science

Dr. Michael J. Person (Course 12, Wallace Astrophysical Observatory)

Professor Kerri Cahoy (Course 16, Course 12)

**Abstract:** Radio interferometry is the current method of choice for deep space astronomy, but in the past few decades optical techniques have become increasingly common. This research seeks to characterize the performance of aperture masking interferometry in the near-infrared at small scales. A mask containing six pairs of apertures at varying diameters and separations was constructed for use with a 24-inch telescope at the MIT Wallace Astrophysical Observatory. Test images of Spica and Jupiter were captured for 28 different telescope configurations, varying aperture separation, aperture diameter, collection wavelength, and exposure time. Lucky imaging was used to account for atmospheric perturbations. Each image was reduced via bias and dark frames to account for sensor noise, and then the full width at half maximum for each image was computed and used as a proxy for maximum angular resolution. The data imply that at small scales aperture size primarily controls the observed maximum angular resolution, but further data are required to substantiate the claim.

## Introduction

Stellar interferometry uses the interference pattern between light collected between two or more apertures to produce a high-resolution image of astronomical phenomena (Born & Wolf, 1959). Interferometric techniques have become increasingly common in the astrophysics and planetary science communities primarily due to their ability to resolve deep-space objects with higher angular resolutions at lower cost than traditional optical methods (Tuthill, 1994). This growing popularity is due to the effective angular resolution of interferometric systems, which is proportional to the diameter between receptors. As a result, receptors spaced far apart and interfered can achieve resolutions far higher than traditional methods could feasibly produce (Hariharan, 2003). Even short baseline systems experience a resolution gain greater than the angular resolution possible with an equally-sized traditional system, making them more efficient at performing the same tasks.

Because of the resolution gains possible using large systems and the ease of interfering long wavelengths to produce high fidelity results, most research regarding stellar interferometry has focused on very large radio telescope arrays, such as the Very Large Array (VLA). As a result, very little work has been done regarding the dynamics of small-scale stellar interferometers. Due to their ability to image higher energy thermal sources such as stars, optical stellar interferometers have become more common in the last twenty years (Tuthill, 1994), but longer wave infrared applications at small scales have remained less well understood, even though they may provide additionally valuable thermal profiles.

Beyond deep space astronomical imaging, stellar interferometry at small scales has potential to be effectively applied to a variety of other applications as well. Amateur astronomers may find stellar interferometry to be significantly more cost effective, as the price for larger equipment grows exponentially. Small subsystems acting as an interferometer may be able to achieve similar results to larger systems at a fraction of the cost. Similarly, the proliferation of small satellites has produced growing interest in portable, inexpensive passive tracking systems for both military and commercial applications. Small-scale stellar interferometry may be effective in both cases, but the optics and the relationship between frequency and performance are not well documented in literature, limiting the development of such systems.

This research proposes to tackle both problems: characterizing the dynamics of small-scale stellar interferometry and understanding the spectral dependence of their performance. To that end, a variable sweep will be performed in simulation to derive a reasonable test matrix of telescope configurations to test. In order to ensure ample light is available to construct high resolution images even at short exposure times, this work will use Spica and Jupiter as the primary targets. Spica is a bright binary star which can be viewed as a point source, while Jupiter is the second brightest object in the night sky after the moon.

## Methods

I – Experimental Setup

All data was collected at the MIT Wallace Astrophysical Observatory in Westford, Massachusetts. The basis of the interferometer was the 24-inch PlaneWave Instruments CDK24 telescope. See Figure 1 for CAD drawings. Data was collected by a ZWO174 Complementary metal–oxide–semiconductor (CMOS) Imager at the rear of the telescope, cooled to 20°C below ambient. See Figure 2 for schematics. A single mask made of an aluminum sheet containing each pair of apertures, depicted in Figure 3. Six geometries were tested: two apertures at 20.4 inches separation for maximum resolution, two apertures at 3.3 inches separation and two apertures at 3.7 inches separation for the zero-fringe condition, as given by $D\_i$ at 750 nm and 850 nm, respectively. The same set is produced for both 1-inch and 2-inch diameter apertures yielding six unique masks, each containing a pair of apertures. Only one pair of apertures are open at a time, with the other five pairs blocked. A cross-section of the apparatus can be seen in Figure 4. Data was also be taken with all apertures covered for the generation of dark and bias frames. See Photometry Section for further discussion on dark and bias frames.

The optimal aperture diameter was found by running over 10,000 simulations sweeping over the variables of collection wavelength, aperture spacing, and aperture diameter. The simulation provided an ideal aperture width of 1 mm. This is to be expected, as the peak of the intensity pattern grows inversely with aperture width for small apertures. To ensure sufficient light is collected to resolve Jupiter and Spica, the minimum aperture diameter is set larger than the optimal value. Beyond 25 mm the intensity peak falls off quickly, but 80% to 90% of the intensity in the diffraction patterns for the 1 mm configuration can be preserved with aperture widths up to that width.

For this simulation, the diffraction pattern of the interferometer was approximated as a double slit with Fraunhofer diffraction. Fraunhofer diffraction assumes parallel wavefronts in the far-field and is valid for F<<1, where F is the Fresnel Number. At larger values of F the wavefronts cannot be assumed to be parallel and are instead represented as parabolic. This regime with parabolic wavefronts in the near-field is called Fresnel diffraction. The Fresnel number is a measure of length-scale and determines what approximations are appropriate for different distances between the double slit and observation point. Because we care about the relative intensity of the diffraction pattern only at the midway point between the two apertures where light is being collected by the CMOS – which is the same whether in the parallel or parabolic wavefront regimes – we can consider this approximation good enough for first pass analysis to determine mask geometries. even though the Fresnel Number of the actual telescope is ~12.5. The expected error given this assumption is not significant enough to alter the outcome of the simulations.

II – Data Collection

Data collection occurred during April of 2019. Spica and Jupiter were chosen as the primary targets. Each mask was bolted to the telescope and the telescope was set to track the target object. With each mask a series of images were taken at two different exposure times for each aperture with each corresponding filter. In total 30,000 images were taken representing 28 different configurations of aperture diameter, aperture separation, wavelength filtering, and exposure time for each target. The full test matrix is listed in Table 1.

III – Photometry and Data Post-Processing

The main goal of post-processing is to reduce blur and atmospheric effects from the data, as well as any other perturbations causing the signal to lose fidelity. It is important to keep in mind physical limitations that cannot be overcome. There are broadly two sources of distortion: atmospheric and telescopic. The atmosphere is a strong absorber in certain bands of the infrared, but there exist "windows" in which transmission is high. The 600 nm to 800 nm band is a prime example of one such window (Sterken and Manfroid, 1992). Taking short exposure images mitigates variable path length and resulting phase shifts due to atmospheric distortions. An additional method of dealing with atmospheric aberrations is knows as *chopping*. Chopping is the process by which consecutive images are taken of the area of the sky of interest and an immediately adjacent area and then the second image is subtracted from the first (Sterken and Manfroid, 1992). This method only works over very short time and spatial scales, however, as atmospheric cells cannot be assumed to be congruent at distances greater than 300 mm (Tuthill,

1994). Factors that influence the noise from the telescope itself include the telescope's black-body radiating temperature and the transmissivity of the CMOS. The transmissivity curve for the filters used on the ZWO174 CMOS can be seen in Figure 5. The i' and z' filters will be used for this experiment, as they provide near complete coverage of the aforementioned 600 nm to 800 nm band (Sterken and Manfroid, 1992). The two filters are centered on 750 nm and 850 nm, respectively, which is the wavelength used for the final simulations. Each filter approximately covers a 100 nm band.

A photometric pipeline is required to reduce raw photo data into a usable image. First, dark, bias, and flat frames must be acquired. A bias frame is the result of taking the fastest possible image with the shutter closed/telescope covered. Doing so eliminates any integration time and measures intrinsic CMOS noise. A dark frame is similar but requires integration times on par with those used to capture the data. Because the CMOS is cooled and the exposure time so short, the dark frame and bias frame can be expected to be very similar for this experiment. In order to construct a master dark and master bias frame, a few hundred of each bias frames and dark frames are captured, and their values averaged pixel-by-pixel to produce a *master bias* and *master dark* frame for that observation session. The flat frame is acquired by imaging a source of uniform illumination to estimate the pixel efficiency of the CMOS. The photometric pipeline is then as follows, as sourced from Poggiani, 2017:

1) The bias frame is subtracted off from the object raw frame.
2) The bias frame is then subtracted from the flat frame, and then that quantity is normalized to its mean value.
3) The first result is then divided by the second to produce a reduced image from the raw photo data.

which can be summarized mathematically as:

$$\text{Reduced Pixel} = \frac{\text{Raw - Bias}}{(\text{Flat - Bias})_\mu}$$

For this experiment, a flat frame was not used and a uniform matrix of 12-bit pixels was used instead. Flat frames primarily correct for variations in the chip, but because the chip is small, has a fairly consistent response, and the angle subtended was small, it was determined that flat frames would not enhance the reduced image quality significantly. Beyond noise from the CMOS, cosmic rays also cause distortions in the form of "hot pixels," or pixel values erroneously high when compared to its surrounding neighbors. These pixels can be averaged with their neighboring pixels to mitigate this effect. This process can be done with a simple filter. The result from this entire process is a reduced image that can be used to make measurements about the relative angular size of targets. An example of a raw image and a reduced image can be found in Figure 6 and Figure 7, respectively.

The final step in the data post-processing pipeline is down sampling. Due to rapidly changing atmospheric conditions, "Lucky Imaging" was used whereby many hundreds of images are taken for a given configuration, and then down sampled for each configuration such that only a small percent of the images are selected for the final dataset. The remaining images represent the best atmospheric conditions for a given night, which mitigates a large proportion of atmospheric noise. The images were selected via a two-step process. First, the maximum valued pixel for each reduced image was found. If that pixel value was greater than 80% of the maximum possible intensity ($2^{12}=4096$ for a 12-bit sensor), then the image was discarded, as pixel values nearing the saturation point of the sensor cause distortions in image quality. Second, the Full Width at Half Maximum (FWHM) is calculated for each of the remaining images. FWHM is twice the distance between the maximum value point and the nearest half maximum point for a data series, and is a measure of how well resolved the data are. Smaller FWHM scores correlate with higher maximum angular resolutions. Because many of the images of Jupiter were overexposed and because Jupiter was too large to constitute a point source, the decision was made to at this time only use data points collected of Spica to characterize the performance of the telescope. An example of one of the overexposed images of Jupiter can be found in Figure 8.

## Results

I – Baseline Calculations

In order to measure the performance of the interferometer, baseline measurements are first required. Taking the 2D Fourier Transform of a given mask configuration yields a point spread function (PSF), which is a reconstruction of the image that would be generated by that mask, as shown in Figure 9. They key difference between a single aperture configuration versus the two-aperture configuration shown is the existence of vertical fringes. These fringes result from the time delay between the signals at each aperture when interfered (Hariharan, 2003). Given the close proximity of the apertures, we expect the fringes to be narrow. Taking the FWHM of the PSF gives the maximum angular resolution of that mask configuration, assuming no atmospheric effects. To demonstrate the resolution gain described previously, the theoretical minimum FWHM for various telescope configurations is provided in Table 2. From the table it is clear to see the resolution gained by the interferometer over the single aperture configuration. Lower values are better because they imply that the system could detect a fainter object.

II – Data Analysis

To determine the effect of the tested parameters on the angular FWHM, each pair of variables was separated and plotted against the demonstrated FWHM. While four variables were varied throughout the experiment, they are not all independent. The aperture separation determines the collection wavelength by setting the zero-fringe condition. The aperture diameter sets the exposure time to prevent oversaturation at larger diameters. Thus, by varying the aperture diameter and separation – the two physical qualities set by the mask – with respect to the achieved FWHM, the full variable space is accounted for. Figure 10 and Figure 11 show these isolated variables. While aperture separation did not appear to be significantly correlated with the demonstrated full width at half maximum, aperture diameter shows strong negative correlation.

## Conclusions

The argument best supported by the data is that variation in full width at half maximum is primarily controlled by the diameter of the aperture. This is intuitive – a larger aperture allows more light to be captured and so allows for the system to more precisely discriminate between targets. However, this result also stands in contrast to the primary justification for using stellar interferometry. Stellar interferometry dominates earth-based deep space imaging primarily because its performance is characterized by the separation of the apertures, not the size of the aperture. The data imply that at small scales, aperture size is of greater importance to the system's resolving power than the separation of the apertures. Therefore, the data suggest that there is a fundamental tradeoff between resolving power and scalability for stellar interferometers. At sufficiently large scales, the separation between the apertures dominates the observed resolution, while at small scales this relationship reverses.

This conclusion has three primary implications. First, there is a hard cutoff to the achievable resolving power of optical stellar interferometers for amateur astronomical use. Given the wavelengths involved, the complexity of interfering optical wavefronts grows exponentially with the baseline, and for the amateur astronomer this growth in complexity will not be worth the marginal gain in resolution, assuming the maximum aperture size is fixed by price limitations. Second, military applications are still viable but remain impractical until a greater understanding of the tradeoff suggested is achieved. Third, commercial applications also remain viable but will be driven primarily by the ability of companies to find profitable uses of the data. The technology alone is not sufficient to commercial adoption.

There exist two primary areas of future work. First, more data are needed to conclusively argue that aperture size controls the observed variation in FWHM. Because so many different configurations were tested, there are not sufficient data points to confirm this trend, especially given its departure from theory. Additionally, more data are required to conclude that aperture separation is as uninfluential in the observed variation in FWHM as the current data suggest. The second area of future work relates to the reversal in trends that the data suggest. If it is the case that at small scales aperture size determines the observed angular resolution, then finding the crossover point, both in terms of minimum aperture size and minimum aperture separation, will be integral to the future design of stellar interferometric systems.

**Acknowledgements**

I would like to first and foremost extend a huge thank you to Professor Lozano, Professor Hall and Jennifer Craig for all their help throughout the year. I would like to thank Professor Cahoy, Dr. Ewan Douglas, and Greg Allen for all their support and for helping me through my ideation. I would like to thank Dr. Michael Person and Tim Brothers of the Wallace Observatory for allowing me to use the observatory as well as meeting with me to discuss my project and teaching me how to use all the equipment. I would additionally like to thank Todd Billings and David Robertson for their help in designing and fabricating the physical mask that allowed this research to happen.

**Figures**

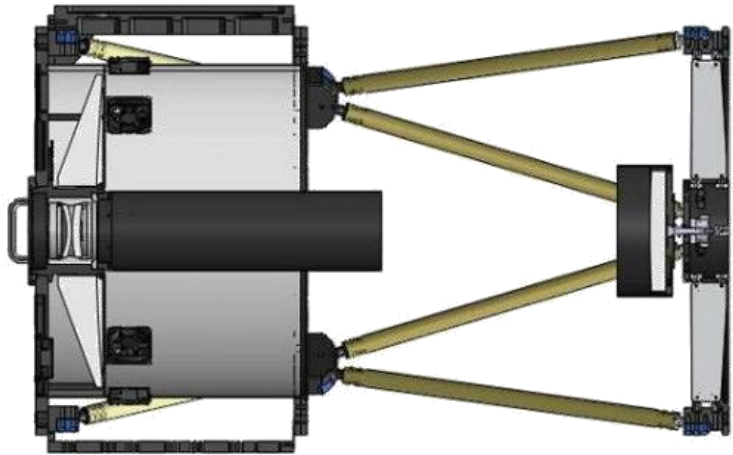

Figure 1: PlaneWave CDK24 Telescope CAD Model. Light enters from the right, is reflected off the primary mirror at the rear, and then bounces off the interior secondary mirror into the CMOS camera.

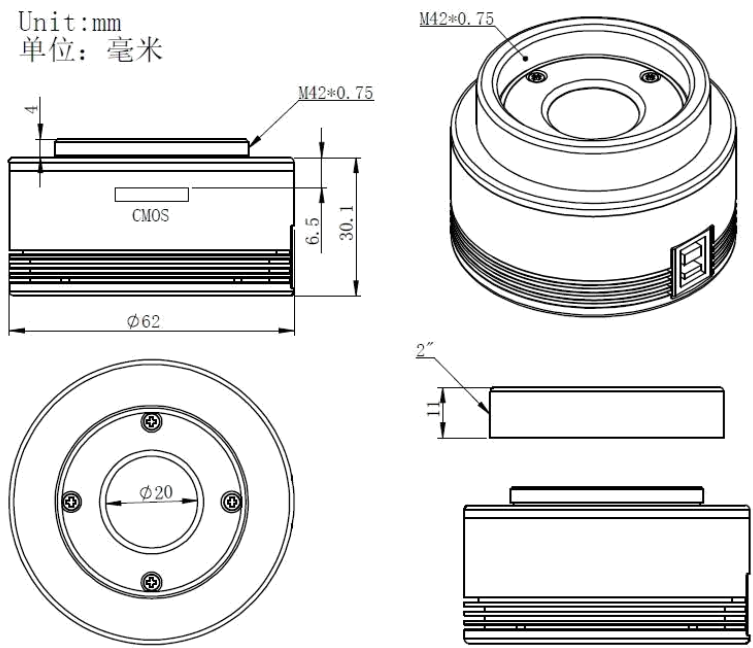

Figure 2: ZWO174 CMOS Schematics. This CMOS is fixed to the back of the telescope (the center on the left in Figure 1).

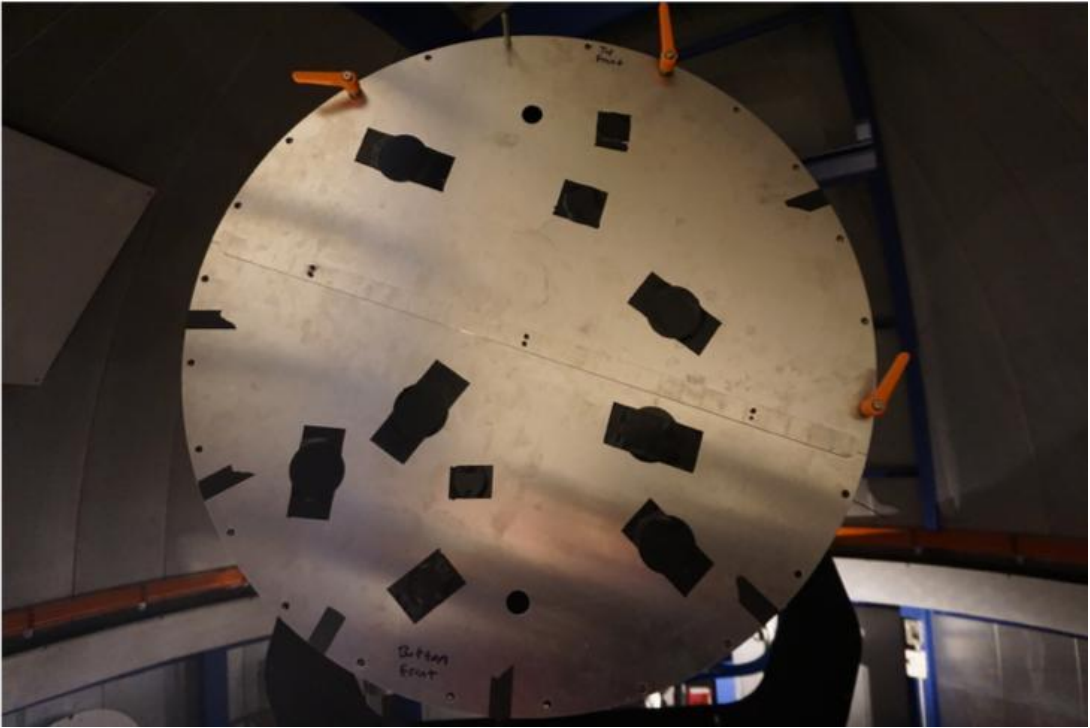

Figure 3: Finished mask bolted to the 24in telescope at the MIT Wallace Observatory. The mask is larger than the 24-inch telescope diameter in order to bolt the mask to the outside casing, as shown in Figure 1. The mask contains 6 pairs of subapertures. There are four centered axial pairs (1L, 1S750, 1S850, 2L) and two off-axial pairs (2S750, 2S850). At the time of imaging the pair of interest is uncovered, while the remaining subaperture array remains covered.

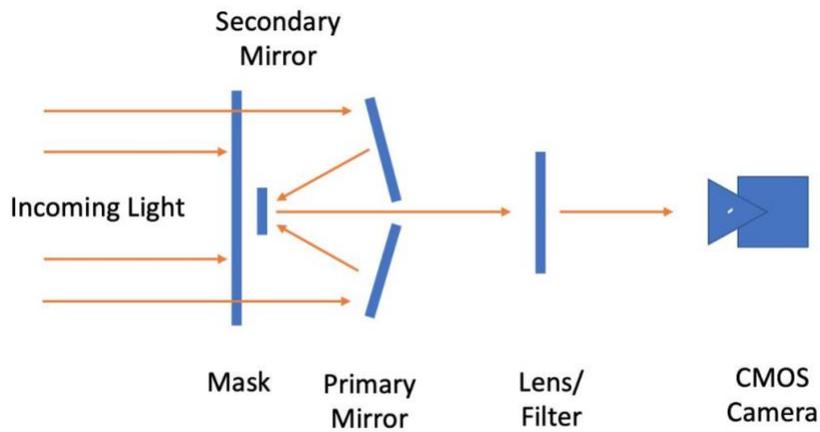

Figure 4: Cross-section of stellar interferometer configuration. Mask configuration and filter are configurable.

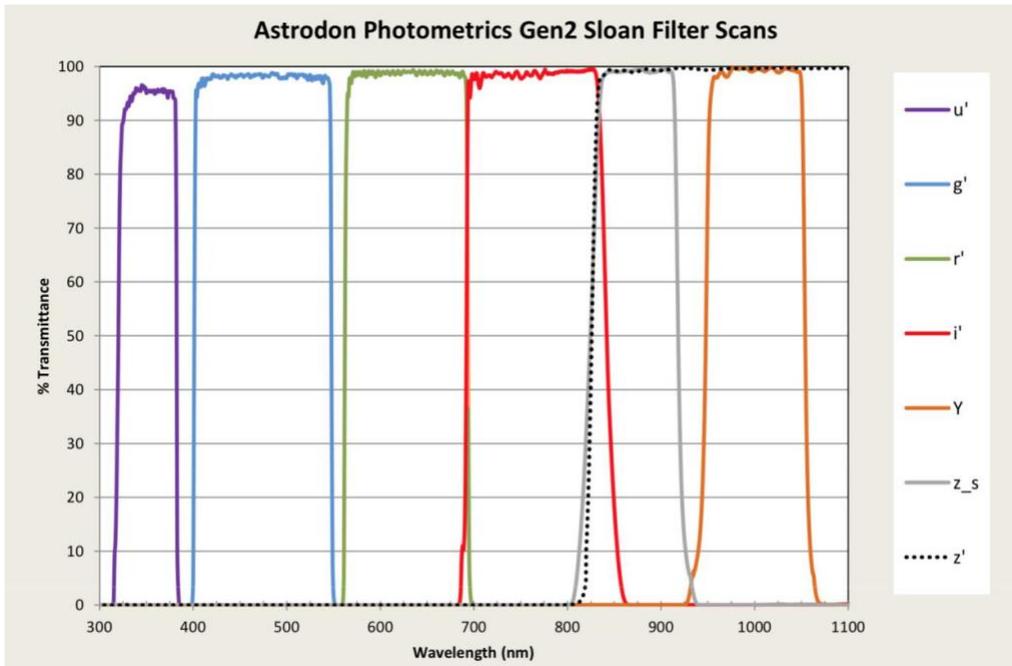

Figure 5: Transmissivity of the ZWO174 CMOS Imager. The bottom axis is wavelength in nm and the right axis is percent transmission. Transmission stays above 98% for the i' and z' filters, which lie in the near-infrared.

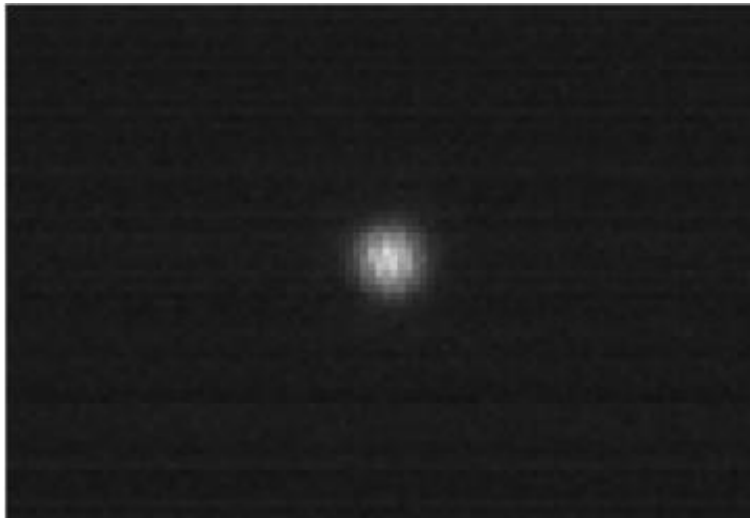

Figure 6: Raw Image of Spica.

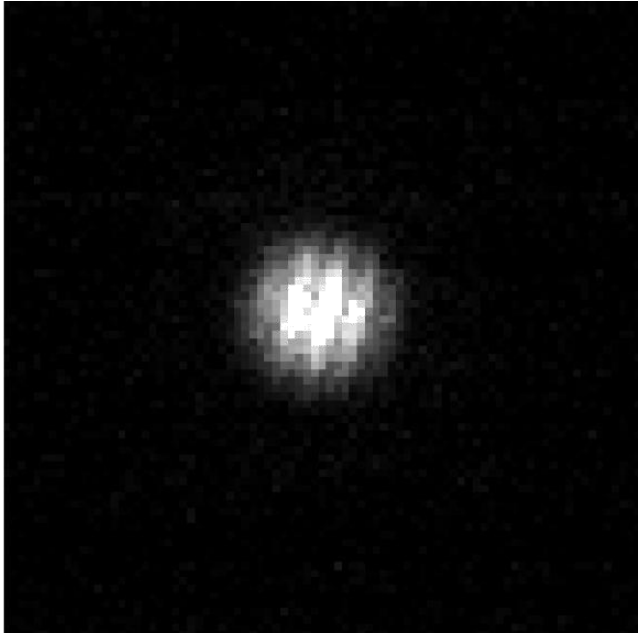

Figure 7: Reduced Image of Spica with clearly exposed fringes.

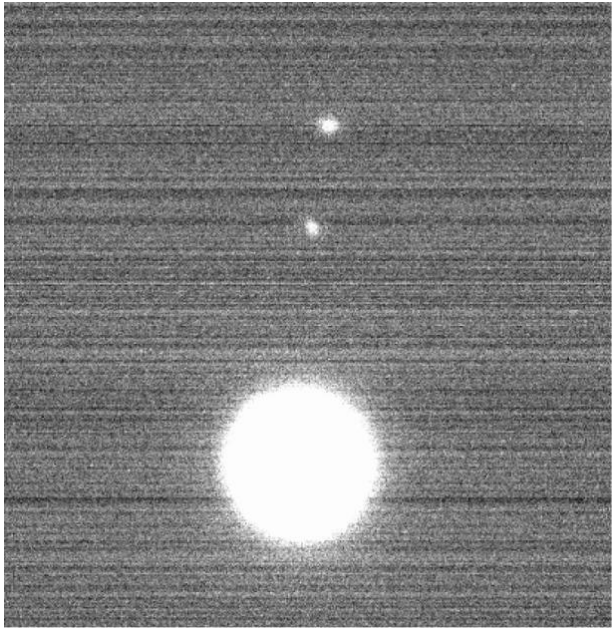

Figure 8: Overexposed raw image of Jupiter and two of its moons.

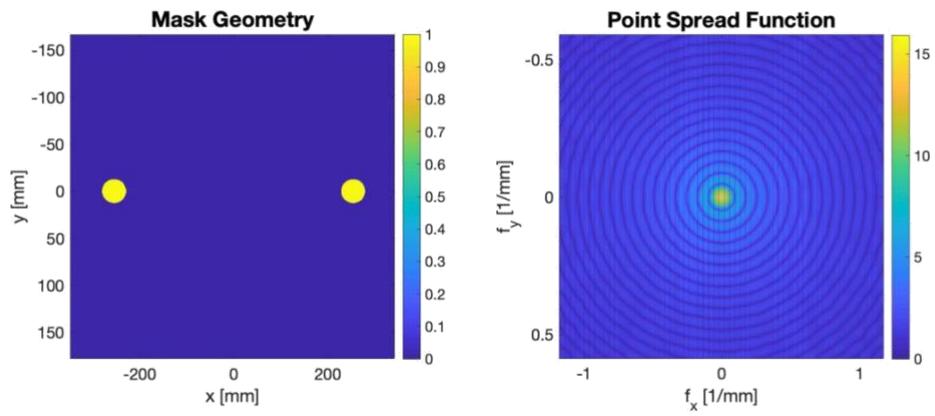

Figure 9: Point spread function (Right) generated from an example mask configuration (Left).

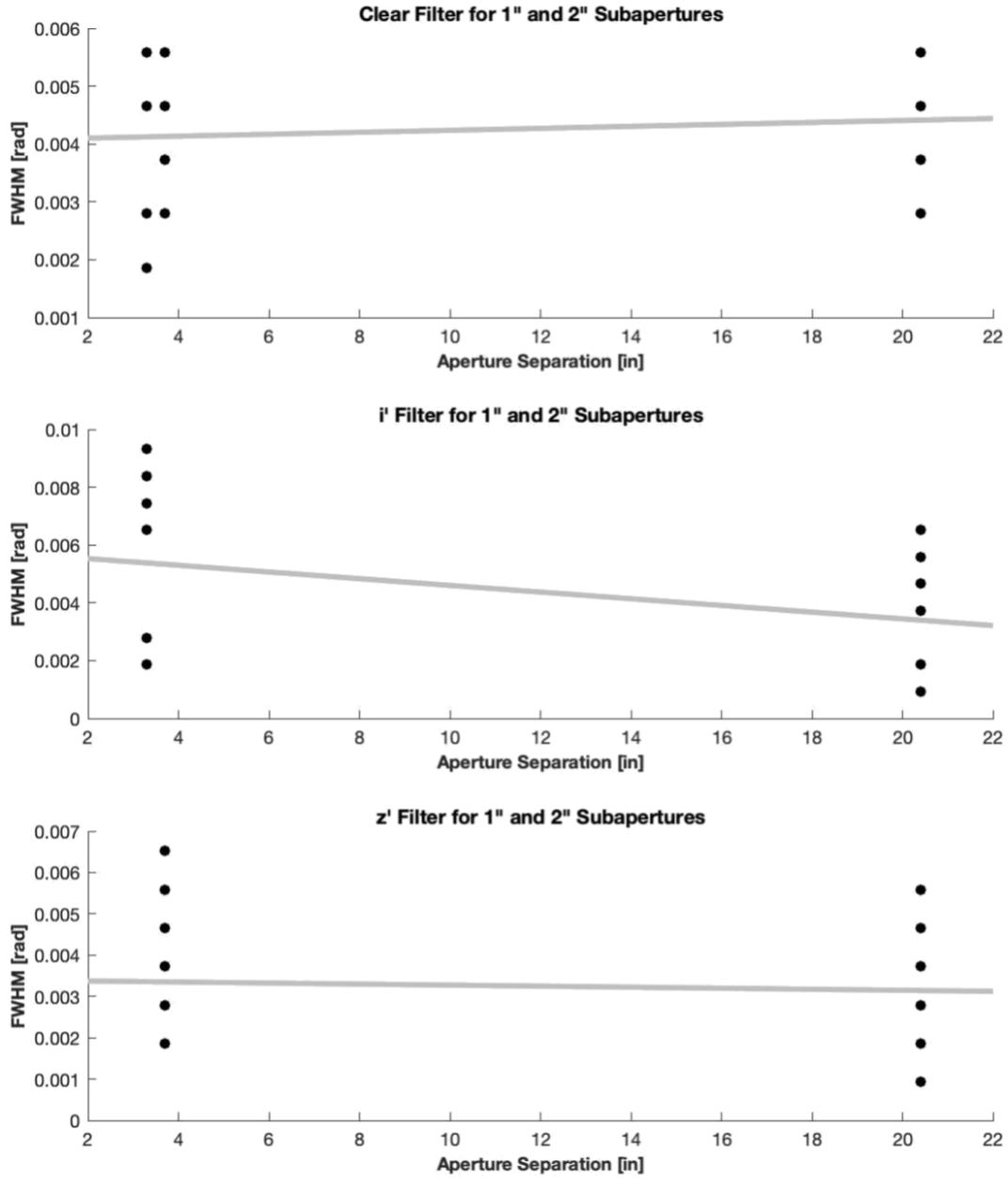

Figure 10: Variation in aperture separation is poorly correlated with FWHM.

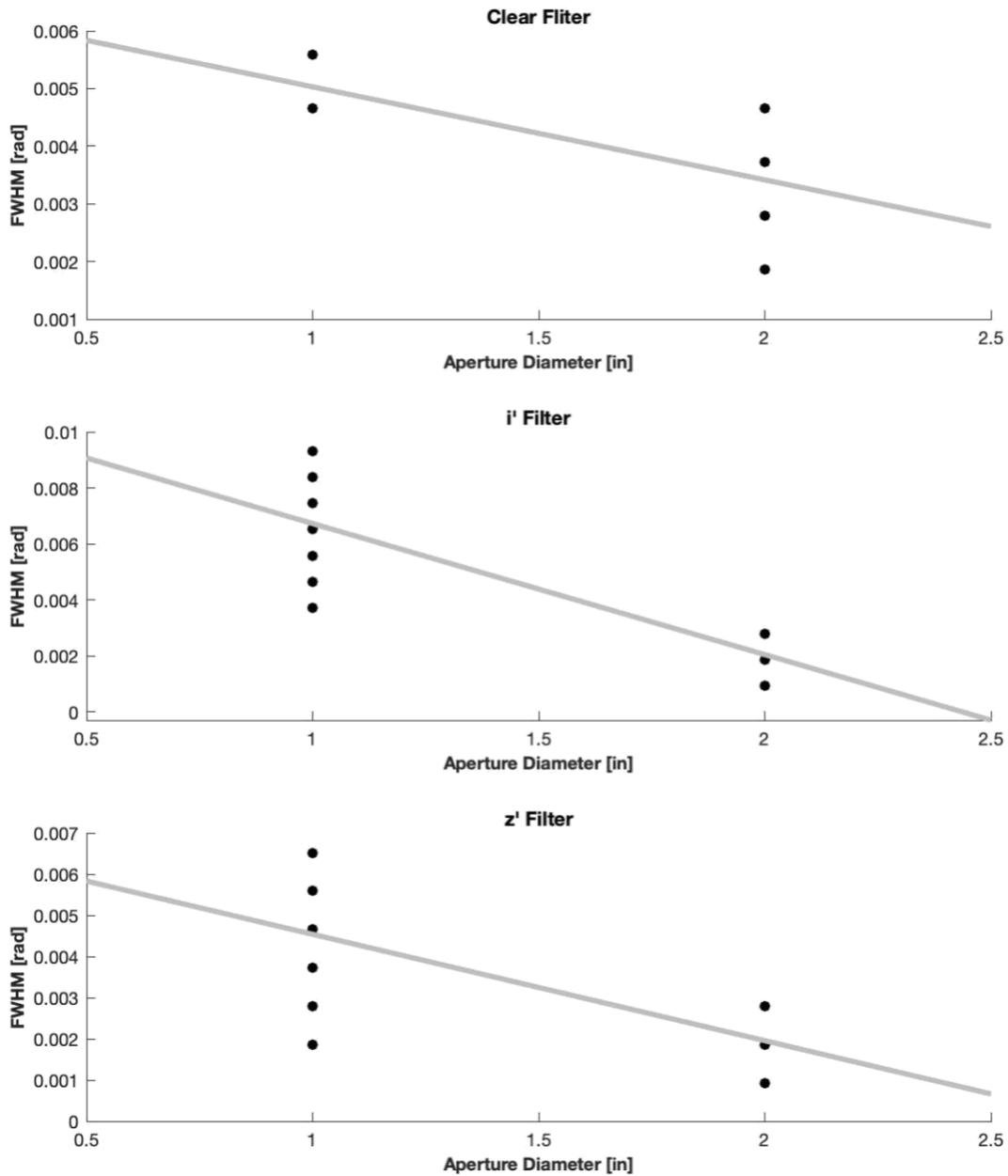

Figure 11: Variation in aperture diameter is negatively correlated with minimizing FWHM.

Table 1 Test Matrix

| Aperture Diameter | Aperture Separation | Filter | Exposure Time |
|---|---|---|---|
| 1 in | 20.4 in | Clear | 100 ms |
| 1 in | 20.4 in | Clear | 5s |
| 1 in | 20.4 in | i' 750 nm | 500 ms |
| 1 in | 20.4 in | i' 750 nm | 5s |
| 1 in | 20.4 in | z' 850 nm | 1250 ms |
| 1 in | 20.4 in | z' 850 nm | 5s |
| 1 in | 3.7 in | Clear | 100 ms |
| 1 in | 3.7 in | Clear | 5s |
| 1 in | 3.7 in | i' 750 nm | 500 ms |
| 1 in | 3.7 in | i' 750 nm | 5s |
| 1 in | 3.3 in | Clear | 100 ms |
| 1 in | 3.3 in | Clear | 5s |
| 1 in | 3.3 in | z' 850 nm | 1250 ms |
| 1 in | 3.3 in | z' 850 nm | 5s |
| 2 in | 20.4 in | Clear | 25 ms |
| 2 in | 20.4 in | Clear | 1250 ms |
| 2 in | 20.4 in | i' 750 nm | 125 ms |
| 2 in | 20.4 in | i' 750 nm | 1250 ms |
| 2 in | 20.4 in | z' 850 nm | 312 ms |
| 2 in | 20.4 in | z' 850 nm | 1250 ms |
| 2 in | 3.7 in | Clear | 25 ms |
| 2 in | 3.7 in | Clear | 1250 ms |
| 2 in | 3.7 in | i' 750 nm | 125 ms |
| 2 in | 3.7 in | i' 750 nm | 1250 ms |
| 2 in | 3.3 in | Clear | 25 ms |
| 2 in | 3.3 in | Clear | 1250 ms |
| 2 in | 3.3 in | z' 850 nm | 312 ms |
| 2 in | 3.3 in | z' 850 nm | 1250 ms |

Table 2 Theoretical Minimum Full Width at Halm Maximum

| Aperture Diameter | Configuration | FWHM |
|---|---|---|
| 1 in | Single Aperture | 0.002 rad |
| 2 in | Single Aperture | 0.0005 rad |
| 20.4 in | Single Aperture | 0.0005 rad |
| 24 in | Single Aperture | 0.000003 rad |
| 1 in | 20.4 in Separation | 0.00009 rad |
| 1 in | 3.7 in Separation | 0.0005 rad |
| 1 in | 3.3 in Separation | 0.0005 rad |
| 2 in | 20.4 in Separation | 0.00004 rad |
| 2 in | 3.7 in Separation | 0.0002 rad |
| 2 in | 3.3 in Separation | 0.0003 rad |